\documentclass[prd,twocolumn,twoside,nofootinbib,preprintnumbers,superscriptaddress,showpacs,showkeys]{revtex4}

\usepackage{amsmath}
\usepackage{amssymb}  
\usepackage{mathrsfs} 
\usepackage{graphicx}
\usepackage{dcolumn}
\usepackage{color}
\usepackage{url}

\def\esl{E \hspace{-0.6em}/\hspace{0.2em}}

\begin{document}

\preprint{IFIC/18-05}

\title{Proton decay and light sterile neutrinos}

\author{Juan C. Helo}
\email{jchelo@userena.cl}
\affiliation{Departamento de F\'{i}sica,
Facultad de Ciencias, Universidad de La Serena,
Avenida Cisternas 1200, La Serena, Chile}
\affiliation{Centro-Cient\'{i}fico-Tecnol\'{o}gico de Valpara\'{i}so,
Casilla 110-V, Valpara\'{i}so, Chile}

\author{Martin Hirsch}
\email{mahirsch@ific.uv.es}
\affiliation{AHEP Group, Instituto de F\'{i}sica Corpuscular
- CSIC/Universitat de Val\`{e}ncia Edificio de Institutos
de Paterna, Apartado 22085, E–46071 Val\`{e}ncia, Spain}

\author{Toshihiko Ota}
\email{tota@yachaytech.edu.ec}
\affiliation{Department of Physics, Yachay Tech,
Hacienda San Jos\'{e} s/n y Proyecto Yachay,
100115 San Miguel de Urcuqu\'{i}, Ecuador}

\begin{abstract}
 
Within the standard model, non-renormalizable operators at dimension six
($d=6$) violate baryon and lepton number by one unit and thus lead to
proton decay. Here, we point out that the proton decay mode with a
charged pion and missing energy can be a characteristic signature of
$d=6$ operators containing a light sterile neutrino, if it is not
accompanied by the standard $\pi^0e^+$ final state. We discuss this
effect first at the level of effective operators and then provide a
concrete model with new physics at the TeV scale, in which the
lightness of the active neutrinos and the stability of the proton are
related.

\end{abstract}



\maketitle

\section{Introduction}

The observed baryon asymmetry of the universe (BAU) requires that
baryon number is violated at high energy scales.  In the standard model
(SM), $B+L$ is
violated by non-perturbative effects, such as instantons
\cite{tHooft:1976rip,tHooft:1976snw} or the sphaleron
\cite{Klinkhamer:1984di}. However, as is well-known, the SM cannot
explain the observed value of the BAU~\cite{Riotto:1999yt}.
Beyond these non-perturbative effects, one can also write down $B$ and
$L$ violating operators at the non-renormalizable level, as has been
discussed already nearly 40 years ago
\cite{Weinberg:1979sa,Wilczek:1979hc,Abbott:1980zj}. Consequently,
many ultra-violet completions of the SM contain $B$ and/or $L$
violating interactions also at the renormalizable level, the prime
example being Grand Unified Theories (GUTs). In particular, $d=6$
operators lead to proton decay, but searches for proton decay so far have
yielded only lower bounds, in the range of ($10^{32}-10^{34}$) yrs,
depending on the final state
\cite{Abe:2013lua,TheSuper-Kamiokande:2017tit,Abe:2014mwa}.  Usually
these negative results are interpreted as a lower limit on the energy
scale of some GUT.

Neutrino masses are much smaller than all other fermion masses.  It is
often argued that this smallness could be understood if neutrinos are
Majorana particles; for a recent review on theoretical aspects of
neutrino masses, see e.g., Ref.~\cite{Cai:2017jrq}.  However, we have
not observed any lepton number violating (LNV) process so far and
limits on neutrinoless double beta decay ($0\nu\beta\beta$) for
example have reached now $10^{26}$ yrs
\cite{KamLAND-Zen:2016pfg,Agostini:2017iyd}. Thus neutrinos could
still be Dirac particles.  Although much less known than the Majorana
case, the study of small Dirac neutrino masses has actually quite a
long history
\cite{Roy:1983be,Chang:1986bp,Mohapatra:1987hh,Mohapatra:1987nx,Balakrishna:1988bn,Babu:1988yq,Rajpoot:1990gy}. Interest
in Dirac neutrino masses has been renewed recently
\cite{Kanemura:2011jj,Farzan:2012sa,Pilaftsis:2012hq,Kanemura:2016ixx,Bonilla:2016diq,Borah:2016zbd,Borah:2016hqn,Wang:2016lve,Ma:2016mwh,Wang:2017mcy,Kanemura:2017haa,Ma:2017kgb,Yao:2017vtm,CentellesChulia:2018gwr},
in particular its possible connection with (cold) dark matter.

In this paper, we ask the question: Could the smallness of the
neutrino mass and the longevity of the proton be related? In other
words, can the mechanism that suppresses proton decay operators also
suppress neutrino masses? First, we study this question at the level
of effective operators.  We point out that the decay mode
$p\rightarrow \pi^{+}$+missing energy ($\pi^++\esl$) with absence of
$p\rightarrow \pi^{0} \ell^{+}$ is a characteristic signature of
effective $d=6$ operators with a light SM singlet fermion, aka
``sterile'' neutrino.  This singlet fermion could be the Dirac partner
of the ordinary neutrinos.
Next, we discuss this in a simple model in which both neutrino
masses and proton decay share a common origin. In the model, $B-L$ is
conserved, thus neutrinos are Dirac particles.  Their masses arise at
the 1-loop level and, as in the model of Ref.~\cite{Farzan:2012sa}, the
particles generating the loop are candidates for the dark matter.
Proton decay arises also at the 1-loop level and shares interactions
and particles with the loop diagram for neutrino masses, therefore
the smallness of neutrino mass is directly related to the longevity of
the proton.
We also consider different experimental constraints on the model
parameters from neutrino masses, proton decay
and searches for lepton flavour violation,
and briefly discuss possible LHC signals of the model.

Before closing this section, we mention that a similar idea, relating
neutrinos and proton decay, has been discussed with a particular model
recently in Ref.~\cite{Gu:2016ghu}. However, differing from our setup, in
Ref.~\cite{Gu:2016ghu} neutrinos are Majorana particles, thus
$0\nu\beta\beta$ decay should exist in their case. We also refer to
Ref.~\cite{deGouvea:2014lva}, where possible relations between
neutrino mass and proton decay are discussed
at the level of higher dimensional effective operators
in the context of GUTs, leading to Majorana neutrinos.
 
The rest of this paper is organized as follows. In the next section we
discuss $d=6$ operators and proton decay modes with a light sterile
neutrino. In section \ref{Sec:Model} we present a concrete model,
which generates $d=6$ proton decay with $\pi^++\esl$ and (Dirac)
neutrino masses at 1-loop and discuss its phenomenology.  We then
close the paper with a short summary.

\section{Effective operators and proton decay modes}
\label{sec:two}

At $d=6$, the baryon-number-violating operators which are invariant
under the SM gauge symmetries can be written as
~\cite{Weinberg:1979sa,Wilczek:1979hc,Abbott:1980zj}:\footnote{%
  Effective operators, which lead to purely leptonic decay modes of
  the proton, appear only at $d=9$, see
  Refs.~\cite{ODonnell:1993kdg,Hambye:2017qix}.  
For proton decay operators with mass dimensions higher than $d=6$, see
also
Refs.~\cite{Babu:2012iv,Lehman:2014jma,Bhattacharya:2015vja,deGouvea:2014lva,Fonseca:2018ehk}.
}
\begin{align}
 \mathcal{O}_{1} =& 
 [\overline{{d_{R}}^{c}}u_{R}] [\overline{Q^{c}}L],
 \label{eq:O1}
 \\
 \mathcal{O}_{2} =&
 [\overline{Q^{c}}Q] [\overline{ {u_{R}}^{c} }e_{R}],
 \label{eq:O2}
 \\
 \mathcal{O}_{3} =&
 [\overline{Q^{c}}Q]_1[\overline{Q^{c}}L]_1,
 \label{eq:O3}
 \\
 \mathcal{O}_{4} =&
 [\overline{Q^{c}}Q]_3[\overline{Q^{c}}L]_3,
 \label{eq:O4}
 \\
 \mathcal{O}_{5} =&
 [\overline{{d_{R}}^{c}}u_{R}][\overline{{u_{R}}^{c}}e_{R}],
 \label{eq:O5}
\end{align}
where for simplicity we have suppressed all flavour and colour
indices.  The subscripts 1 and 3 at the brackets represent the singlet
and the triplet combinations of $SU(2)_{L}$.
Operators with the same fields but Lorentz structures different from
$\mathcal{O}_{1\text{-}5}$ can be rewritten as combinations of these
basis operators via Fierz transformations. For example,
\begin{align}
 [\overline{Q^{c}}(\sigma^{\rho})u_{R} ] [\overline{{d_{R}}^{c}}
 (\overline{\sigma}_{\rho})L] = 2\mathcal{O}_{1}.
 \nonumber
\end{align}
All effective operators listed above respect $B-L$, but have
$\Delta(B+L)=2$.

Extending the particle content of the SM by singlet fermion fields
$N$ with one unit of lepton number, one can write down the
following two additional operators, which are both invariant under SM
gauge transformations and $B-L$,
see e.g.,
Ref.~\cite{delAguila:2008ir,Alonso:2014zka,Liao:2016qyd,Merlo:2016prs}\footnote{
Nucleon decays with the operator $\mathcal{O}_{N2}$ 
are discussed in Ref.~\cite{Davoudiasl:2014gfa}
in relation with dark matter.
}
\begin{align}
 \mathcal{O}_{N1}
 =&
 [\overline{Q^{c}} Q][\overline{{d_{R}}^{c}} N],
 \\
 \mathcal{O}_{N2} =&
 [\overline{{u_{R}}^{c}}d_{R}] [\overline{{d_{R}}^{c}}N ].
\end{align}
%
\begin{table}[t]
 \begin{tabular}{cccc}
  \hline \hline
  Modes $(p)$ & $\pi^{+}+\esl$ & $\pi^{0}e^{+}$ & $K^{+}+\esl$
	      \\
  \hline
  Current [yrs]
  & $3.9 \cdot 10^{32}$~\cite{Abe:2013lua}
      & $1.6\cdot 10^{34}$~\cite{TheSuper-Kamiokande:2017tit}
	  & $5.9 \cdot 10^{33}$~\cite{Abe:2014mwa}
	      \\
  Future
  [yrs]
  &
      & $1.2 \cdot 10^{35}$~\cite{Hyper-Kamiokande:2016dsw}
	  & $ >3 \cdot 10^{34}$~\cite{Acciarri:2015uup}
      \\
  \hline
  $\mathcal{O}_{1}$
  &
      $\checkmark$ &  $\checkmark$ &  $\checkmark$  
	      \\
  $\mathcal{O}_{2}$
  &
      --- & $\checkmark$ & ---
      \\
  $\mathcal{O}_{3}$
  &
      $\checkmark$ & $\checkmark$ & $\checkmark$
      \\
  $\mathcal{O}_{4}$
  &
      --- & --- & $\checkmark$
	      \\
  $\mathcal{O}_{5}$
  & --- & $\checkmark$ & ---
      \\
  \hline
  $\mathcal{O}_{N1}$
  & $\checkmark$ & --- & $\checkmark$
      \\
  $\mathcal{O}_{N2}$
  & $\checkmark$ & --- & $\checkmark$
  \\
  \hline \hline
  \end{tabular}
 \caption{Operators and proton decay modes. The numbers in the
   ``Current'' and the ``Future'' rows are the current bounds and the
   future sensitivities at 90 \% C.L. Only the operators
   $\mathcal{O}_{N1}$ and $\mathcal{O}_{N2}$ generate $\pi^++\esl$
   (missing energy), without producing the decay $\pi^0 e^+$.}
 \label{Tab:Modes}
\end{table}

Proton decay final states differ depending on the operator under
consideration.  Operators and the corresponding decay modes are
summarized in Tab.~\ref{Tab:Modes}, together with the current bounds
and future
sensitivities~\cite{Abe:2013lua,TheSuper-Kamiokande:2017tit,Abe:2014mwa,Hyper-Kamiokande:2016dsw,Acciarri:2015uup}.
The final state with a charged pion and missing energy can be
generated by $\mathcal{O}_{1,3}$ and $\mathcal{O}_{N1,N2}$.  For the
case of $\mathcal{O}_{1,3}$ the final state $\pi^++\esl$ is caused by
the emission of a left-handed neutrino. Isospin symmetry then tells us
that $\mathcal{O}_{1,3}$ also generate the process with
the corresponding left-handed charged lepton, which is
$p \rightarrow \pi^{0} \ell^{+}$.
The decay rates of the two $SU(2)_{L}$-related processes are expected to
fulfill the following ratio
(cf. e.g., Refs.~\cite{Senjanovic:2009kr,Golowich:1981sb}):
\begin{align}
 \Gamma(p \xrightarrow[]{\mathcal{O}_{1,3}} \pi^{+} \bar{\nu}_{e})
 =
2 \Gamma(p \xrightarrow[]{\mathcal{O}_{1,3}} \pi^{0} e^{+}). 
\end{align}
Conversely, the operators $\mathcal{O}_{N1,N2}$ do not have a charged
lepton counter part and thus cannot generate the decay mode with a
neutral pion and a charged lepton. It seems natural then to suppose
that the discovery of proton decay with final state $p \rightarrow
\pi^{+}+\esl$ with simultaneous absence of the $\pi^{0} \ell^{+}$ mode
suggests that the process is caused by an operator containing a SM
singlet fermion $N$. Since the decay mode $p \rightarrow \pi^{0} e^{+}$ is
more strongly constrained than $p \rightarrow \pi^{+}+\esl$, a
discovery of the $\pi^{+}$+missing mode in the next round of proton
decay searches would therefore hint at the existence of a light
sterile neutrino with mass below $m_p - m_{\pi}$.

The effective operators $\mathcal{O}_{N1,N2}$ with a sterile neutrino
also generate neutron decay process with $\pi^{0}+\esl$. One expects
them to follow a particular ratio:
\begin{align}
 \Gamma(p \xrightarrow[]{\mathcal{O}_{N1,N2}} \pi^{+} \bar{N})
 =
 2 \Gamma (n \xrightarrow[]{\mathcal{O}_{N1,N2}} \pi^{0} \bar{N}).
\end{align}
The operators and the corresponding neutron decay modes are listed
in Tab.~\ref{Tab:Neutron}, and the current bounds and the future
sensitivities are found in
Refs.~\cite{Abe:2013lua,TheSuper-Kamiokande:2017tit,Hyper-Kamiokande:2016dsw}.
Again, for $\mathcal{O}_{N1,N2}$ there are no decays to charged 
leptons, $\pi^{-}e^{+}$, thus the same logic holds also for neutron 
decays and observation of $\pi^{0}+\esl$ can be interpreted as a 
hint for a light sterile state.

\begin{table}[t]
 \begin{tabular}{ccc}
  \hline \hline
  Modes $(n)$ & $\pi^{0}+\esl$ & $\pi^{-}e^{+}$ 
	      \\
  \hline
  Current [yrs]
  & $1.1\cdot10^{33}$~\cite{Abe:2013lua}
      & $5.3 \cdot 10^{33}$~\cite{TheSuper-Kamiokande:2017tit}
	      \\
  Future
  [yrs]
  &
      & $3.8\cdot 10^{34}$~\cite{Hyper-Kamiokande:2016dsw}
	  
      \\
  \hline
  $\mathcal{O}_{1}$
  & $\checkmark$
      & $\checkmark$
	      \\
  $\mathcal{O}_{2}$
  & ---
      & $\checkmark$
      \\
  $\mathcal{O}_{3}$
  & $\checkmark$
      & $\checkmark$
      \\
  $\mathcal{O}_{4}$
  & ---
      & ---
	      \\
  $\mathcal{O}_{5}$
  & ---
  & $\checkmark$
      \\
  \hline
  $\mathcal{O}_{N1}$
  & $\checkmark$ & ---
      \\
  $\mathcal{O}_{N2}$
  & $\checkmark$ & --- 
  \\
  \hline \hline
  \end{tabular}
 \caption{Same as Tab.~\ref{Tab:Modes} but for neutron decay modes.
   Here again, the discovery of $\pi^{0}+\esl$ with simultaneous
   absence of $\pi^{-}\ell^{+}$ suggests the existence of an effective
   operator containing a light sterile neutrino.}
 \label{Tab:Neutron}
\end{table}

At this point, we have to add a word of caution to the above
discussion. The simple arguments presented are based on $SU(2)_L$
invariance, used in the construction of all non-renormalizable
operators. While certainly $SU(2)_L$ is restored at high energies,
and thus, all ultra-violet completions of the SM should respect it,
this by no means implies that $SU(2)_L$ breaking effects are guaranteed
to be negligible. 
We will discuss briefly two particular examples for setups
with possibly sizable $SU(2)_L$ violating effects.

The study of proton decay in supersymmetric (SUSY) GUTs has a long
history, see e.g.,
Refs.~\cite{Langacker:1980js,Sakai:1981pk,Dimopoulos:1981dw,Ellis:1981tv,Nath:1985ub,Hisano:1992jj,Lucas:1996bc,Goto:1998qg,Murayama:2001ur,Hisano:2013exa}.
In SUSY-GUT frameworks, the leading contributions to proton decay come
usually from one-loop diagrams which contain $B$ and $L$ violating
dimension-five operators with two sfermions and the so called ``dressing''
of the operators with a gaugino or a higgsino, which converts the
sfermions to the corresponding fermions.
The flavour structure of the Yukawa interactions entering
in this ``dressing'' diagram can lead to a large
difference between the rate of the decay $p \rightarrow \pi^{+}
\bar{\nu}$ and that of $p \rightarrow \pi^{0}\ell^{+}$.
In fact, the $\pi^{+}\bar{\nu}$ decay can become more important than the
$\pi^{0}\ell^{+}$ mode in a large class of the SUSY-GUT models, see
for example Ref.~\cite{Nath:1985ub}.
However, the dominant proton decay mode in SUSY-GUTs is
in general $p\rightarrow K^{+}\bar{\nu}$.
Therefore, the discovery of the $p \rightarrow
\pi^{+}+\esl$ final state with absence of the $\pi^{0}\ell^{+}$
{\em and} $K^{+}+\esl$ final states, could still be interpreted
as a hint for a light sterile state even in supersymmetric frameworks.
A notable exception from this argument is, however, the SUSY $SO(10)$
model discussed in Ref.~\cite{Goh:2003nv}. Here, the $p \rightarrow
\pi^{+} \bar{\nu}$ may become the dominant mode in part of the
parameter space, in which the decay $p\rightarrow
K^{+}\bar{\nu}$ is minimized in order to obey the experimental bounds.

As the second example for the $SU(2)_{L}$ violation effect,
we mention the model of Ref.~\cite{Gu:2016ghu}. Here,
proton decay is generated by a $d=7$ operator $(u_X d_X)d_R(
H^{0*} {\bar \nu} - H^-l^+)$, where $X=L$ or $X=R$ and $H$
stands for the SM Higgs field.\footnote{ This $d=7$ operator violates $B-L$ as is demonstrated in Ref.~\cite{Kobach:2016ami}. }
The vacuum expectation value of the Higgs field 
picks out the neutrino term exclusively
from the effective operator.
The $\pi^{0}\ell^{+}$ mode can also be generated
from the $d=7$ operator,
but it is suppressed relative to the $\pi^+ +\esl$ mode,
because it requires an extra $W$ insertion.
Thus, with $d=7$ operators (and correspondingly
higher dimensional ones) involving Higgs fields, $SU(2)_L$ violation 
occurs naturally, restricting our argument to $d=6$ operators.

In short, we have pointed out that (in the absence of any experimental
indication for TeV-scale supersymmetry) the combination of different
proton decay final states can provide hints for (or against) the
existence of light sterile neutrino states. This argument is based on
the assumption of (at least approximate) $SU(2)_L$ invariance.  We note
that all $d=6$ operators conserve $B-L$, thus these sterile states
could be the Dirac partners of the ordinary neutrinos.  In the next
section, we will discuss the relation between the stability of proton
and the lightness of the neutrino in a concrete model with TeV-scale
new physics (NP).

\section{Longevity of proton and lightness of neutrino}
\label{Sec:Model}

Here we discuss how a possible relation between the smallness of
neutrino masses and the stability of the proton can arise in a
concrete model. The particle content of this model is given in
Tab.~\ref{Tab:Model}.
%
\begin{table}[t]
 \begin{tabular}{ccc}
  \hline \hline
  $(SU(3)_{c},SU(2)_{L}, U(1)_{Y}, U(1)_{B-L})$
  & $Z_{2}^{(A)}$ & $Z_{2}^{(B)}$
	  \\
  \hline
  $L({\bf 1},{\bf 2},-1/2,-1)$ & $+$ & $-$
	  \\
  $e_R({\bf 1},{\bf 1},-1,-1)$ & $+$ & $-$
	  \\
  Other SM particles & $+$ & $+$
	  \\
  \hline
  $N({\bf 1},{\bf 1}, 0, -1)$ & $-$ & $+$
	  \\
  \hline
  $\eta({\bf 1},{\bf 2}, +1/2,0)$ & $-$ & $-$
	  \\
  $S'({\bf 1},{\bf 1}, 0,0)$ & $-$ & $+$
	  \\
  $\psi'=(\psi'_{L},\psi'_{R}) ({\bf 1},{\bf 1}, 0,-1)$
  & $(+,-)$ & $(+,+)$
	  \\
  $S({\bf 3},{\bf 1},-1/3,-2/3)$ & $-$ & $+$
	  \\
  $\psi=(\psi_{L},\psi_{R})(\overline{\bf 3},{\bf 2},-1/6,-1/3)$
  & $(-,-)$ & $(+,+)$
	  \\
  \hline \hline
 \end{tabular}
 \caption{Particle content of the model, together with the
   corresponding charge assignments. $Z_{2}^{(A)}$ is broken only by
   the mass term of the fermion field $\psi'$. $Z_{2}^{(B)}$ is broken
   softly via the $\mu$-term, see Eq. (\ref{eq:Lagrangian}).}
 \label{Tab:Model}
\end{table}
%
The model is described by the following Lagrangian:\footnote{%
For a complex scalar $S'$, one can write
  also the terms $(Y'_{2})^{\alpha} (\overline{{\psi_{L}}^{c}})
  (Q_{\alpha}) S'$ and
  $(Y'_{3})^{\alpha}(\overline{\psi_{L}'})(N_{\alpha})
  S'^{\dagger}$. Since these do not lead to any new phenomenology we
  have suppressed them for brevity.}
\begin{equation}\label{eq:lagAB}
 \mathscr{L} = \mathscr{L}_1 + \mathscr{L}_2
\end{equation}
where
\begin{align}
 \mathscr{L}_1 =&
 (Y_{1})^{\alpha}
 (\overline{Q^{c}}_{\alpha})({\psi_{R}}^{c})S
 +
 (Y_{2})^{\alpha}
 (\overline{{\psi_{L}}^{c}}) (Q_{\alpha}) S'^{\dagger}
 \nonumber
 \\
 &+ (Y_{3})^{\alpha}(\overline{\psi_{L}'})(N_{\alpha}) S'
 + (Y_{4})^{\alpha}(\overline{{d_{R}}^{c}}_{\alpha})(\psi'_{R})S^{\dagger}
 \nonumber
 \\
 &+(Y_{\nu})^{\alpha}(\overline{\psi'_{R}})(L_{\alpha})\eta
 - M_{\psi} (\overline{\psi_{L}})(\psi_{R})
 - M_{\psi'} (\overline{\psi'_{L}})(\psi'_{R})
 \nonumber
 \\
 & + \mu \eta^{\dagger} H S'
 +
 {\rm H.c.}
 - M_{S}^{2} S^{\dagger } S
 - M_{S'}^{2} S'^{\dagger} S',
 \label{eq:Lagrangian}
\end{align}
and
\begin{align}
 \mathscr{L}_2 =&
 (Y_{4}')^{\alpha} (\overline{{d_{R}}^{c}}_{\alpha}) N S^{\dagger}
 +
 {(Y_N)_{\alpha}}^{\beta} {\overline N}^{\alpha} L_{\beta}\eta
 \nonumber
 \\
 &- M_{\psi'N} \overline{\psi'_{L}} N
 + {\rm H.c.}.
 \label{eq:Lagrangian2}
\end{align}
The terms in Eq.~(\ref{eq:Lagrangian}) are needed for the consistency 
of the model, while the terms in Eq.~(\ref{eq:Lagrangian2}) are 
optional, see below. 
Here the parameter $\mu$ has dimension of mass. We have assumed
that there are three copies of $N$, but only one copy of the fields 
$\psi_{L}$, $\psi_{R}$, $\psi'_{L}$ and $\psi'_{R}$, for 
simplicity. Note that at least two copies of $N$ are needed, 
since neutrino oscillation data require two non-zero 
mass eigenstates for the active neutrinos. 

\begin{figure*}[ht]
 \unitlength=1cm
 \hspace{-1cm}
 \begin{picture}(6,3.5)
  \put(0,0.5){\includegraphics[width=6cm]{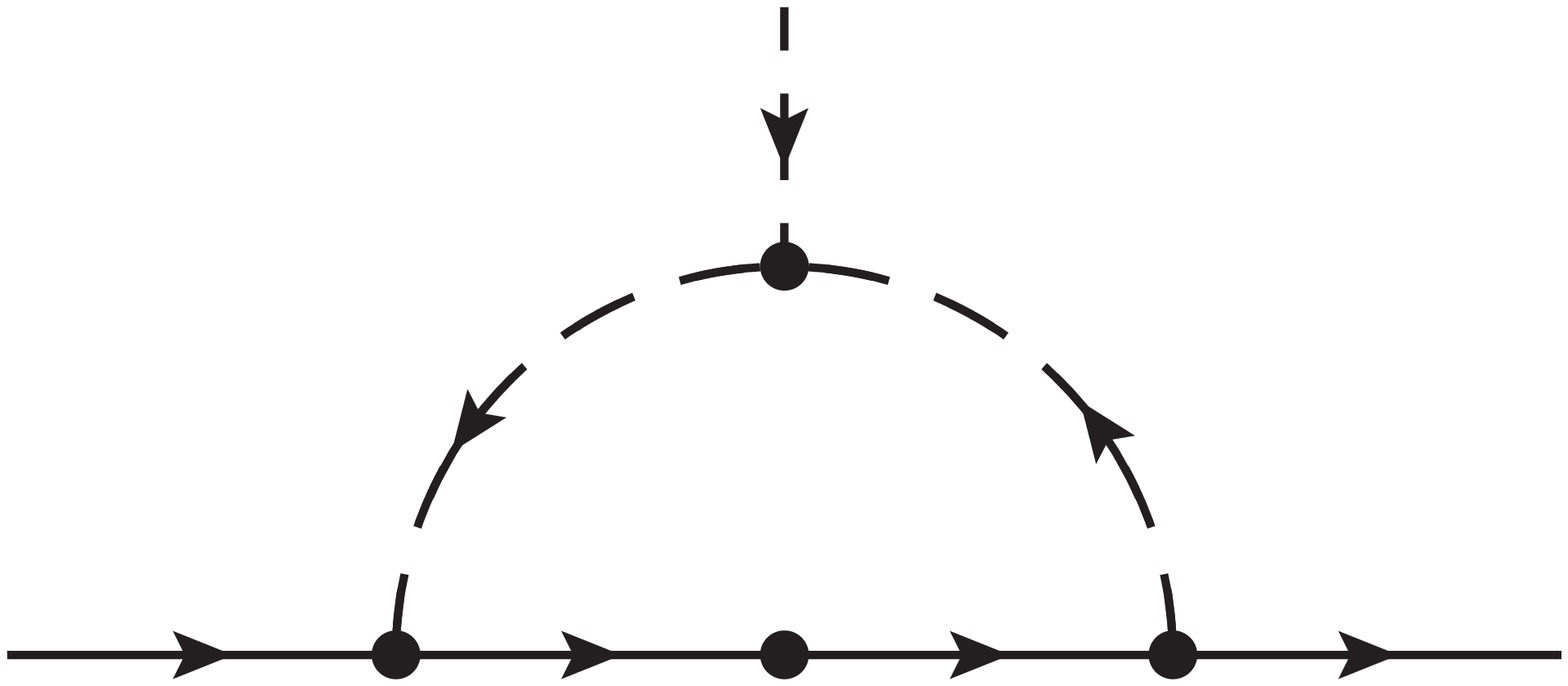}}
  \put(-0.2,0.2){$L(+,-)$}
  \put(1.2,0.9){$Y_{\nu}$}
  \put(1.6,0.2){$\psi'_{R}(-,+)$}
  \put(3.1,0.2){$\psi'_{L}(+,+)$}
  \put(2.8,0.9){$M_{\psi'}$}
  \put(4.5,0.9){$Y_{3}^{\dagger}$}
  \put(5.1,0.2){$N(-,+)$}
  \put(4,1.8){$S'(-,+)$}
  \put(0.8,1.8){$\eta(-,-)$}
  \put(2.85,3.1){$H(+,+)$}
  \put(2.85,1.7){$\mu$}
 \end{picture}
 \hspace{1.5cm}
 \begin{picture}(5.5,3.5)
  \put(0.4,0.45){\includegraphics[width=5.5cm]{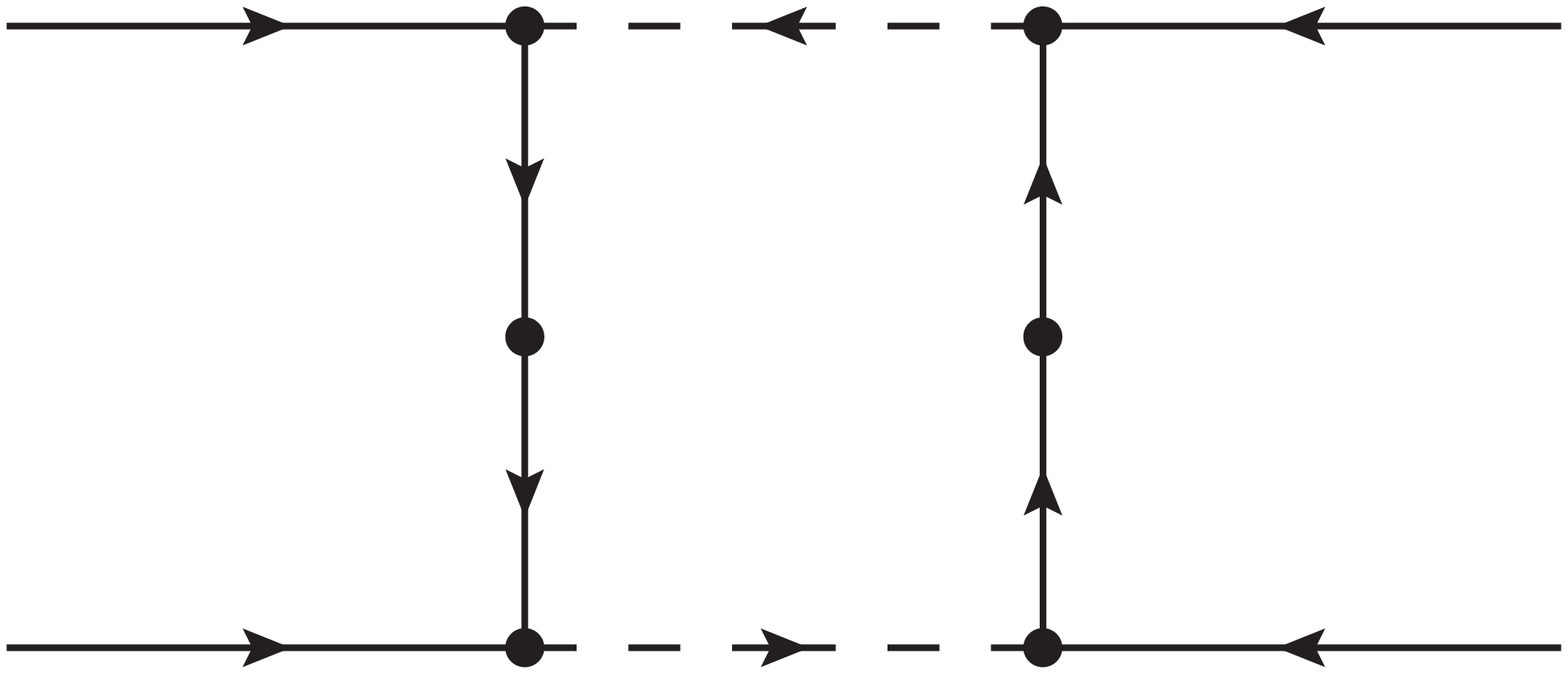}}
  \put(0,3){$Q_{\alpha}(+,+)$}
  \put(0,0.1){$Q_{\beta}(+,+)$}
  \put(5.5,0.1){$N_{\delta}(-,+)$}
  \put(5.5,3){$d_{R \gamma}(+,+)$}
  \put(0.85,2.1){$\psi_{R}(-,+)$}
  \put(0.85,1){$\psi_{L}(-,+)$}
  \put(2.45,1.5){$M_{\psi}$}
  \put(2.55,3){$S(-,+)$}
  \put(4.2,1){$\psi'_{L}(+,+)$}
  \put(4.2,2.1){$\psi'_{R}(-,+)$}
  \put(3.35,1.5){$M_{\psi'}$}
  \put(2.55,0.1){$S'(-,+)$}
  \put(2.1,2.9){$Y_{1}$}
  \put(2.1,0.2){$Y_{2}$}
  \put(3.9,0.2){$Y_{3}$}
  \put(3.9,2.9){$Y_{4}$}
 \end{picture}
 \caption{Diagrams for the Dirac
 neutrino mass and the $d=6$ proton decay operator.  The signs
 $+$/$-$ represent the charges under the $(Z_{2}^{(A)},Z_{2}^{(B)})$
 symmetries.
 $Z_{2}^{(A)}$ is broken only via the mass term $M_{\psi'}$ of the fermion
 field $\psi'$, and $Z_{2}^{(B)}$ is softly broken with the trilinear
 $\mu$ term. See text.}
 \label{Fig:Diagrams}
\end{figure*}

The transformation properties/charges of all new particles are listed
in Tab.~\ref{Tab:Model}.   Both, $\eta$ and $S'$
  have to be inert scalars, i.e. $\langle \eta^0\rangle \equiv 0
  \equiv \langle S'\rangle $.  Two additional $Z_{2}$ symmetries are
imposed:  Since the Dirac neutrino mass term and
  the proton decay operator $\mathcal{O}_{N1}$ violate the
  $Z_{2}^{(A)}$ symmetry, they can appear only in connection with the
  mass term of $\psi'$, which guarantees that they are radiatively
  generated. The loop diagrams for the proton decay operator and the
  Dirac neutrino mass term are shown in Fig.~\ref{Fig:Diagrams}.  The
  symmetry $Z_{2}^{(B)}$ is necessary to eliminate unwanted terms in
  the Lagrangian that could generate the standard proton decay
  operators, see Eqs.~(\ref{eq:O1})-(\ref{eq:O5}).
   Note that, since
  $Z_{2}^{(B)}$ is broken softly by the $\mu$-term, the proton decay
  operators $\mathcal{O}_{1}$-$\mathcal{O}_{5}$ will appear
  at higher loop order or as higher dimensional diagrams.
  We have found that for these operators the lowest
  order contribution comes from  a 1-loop $d=8$ diagram which
  gives only sub-dominant effects in comparison with
  the diagram shown in Fig.~\ref{Fig:Diagrams}.

In addition to the neutrino mass diagram on the left of 
Fig.~\ref{Fig:Diagrams}, there is also a possible diagram 
with $N$ and $\psi'_R$ exchanged, due to the terms in
Eq.~(\ref{eq:Lagrangian2}). The presence of these terms allows, 
in principle, to suppress the neutrino mass additionally 
by a Dirac seesaw \cite{Roy:1983be}. However, these terms 
can equally well be eliminated by an additional $Z_2$ for the 
particles in the loops in Fig.~\ref{Fig:Diagrams}. This option 
has the advantage that the model at the same time could also 
explain the dark matter problem, see below. We will therefore not 
discuss the possibility of obtaining a Dirac seesaw suppression 
further.
The additional Yukawa interaction $Y_{4}'$ provides
another loop diagram similar to the one shown on
the right side in Fig.~\ref{Fig:Diagrams}.
However, since this contribution does not make any qualitative
difference in our argument,
we do not discuss this interaction further
in the present study.

%
%
The model described by Eq.~\eqref{eq:Lagrangian} and
Tab.~\ref{Tab:Model} draws some inspiration from the scotogenic model
for Dirac neutrino mass, which was presented in Ref.~\cite{Farzan:2012sa}.
However, the authors of Ref.~\cite{Farzan:2012sa} discuss only
Dirac neutrino masses, while our variant induces also proton decay
and in fact, relates the rates of the two processes as follows.
In our model the charge assignments of one of the $Z_{2}$ symmetries
are modified to allow the proton decay operator. More concretely,
the $Z_{2}^{(A)}$ symmetry is broken with the mass term of the new fermion
field $\psi'$ in our choice, instead of the trilinear scalar coupling
$\mu$. With this assignment, both the Dirac neutrino mass and
the proton decay can occur only via a $Z_2^{(A)}$ symmetry breaking
mass term $M_{\psi'} (\overline{\psi'_{L}})(\psi'_{R})$.

Let us give a rough estimate for the resulting proton lifetime
and Dirac neutrino mass.  The Yukawa interactions given in
Eq.~\eqref{eq:Lagrangian} mediate the effective operator
$\mathcal{O}_{N1}$, and the coefficient can be evaluated as
\begin{align}
 \mathscr{L}_{\text{eff}}
 =
 (Y_{1})^{\alpha}
 (Y_{2})^{\beta}
 (Y_{3})^{\delta}
 (Y_{4})^{\gamma}
 M_{\psi} M_{\psi'}
 I_{4}
 \mathcal{O}_{N1},
\end{align}
where $I_{4}$ is the loop integral function defined as
\begin{align}
 I_{4}
 \equiv&
 \int
 \frac{{\rm d}^{d}k}{(2\pi)^{d}{\rm i}}
 \frac{1}{
 [k^{2} - M_{S}^{2}]
 [k^{2} - M_{S'}^{2}]
 [k^{2} - M_{\psi}^{2}]
 [k^{2} - M_{\psi'}^{2}]
 } .
\end{align}
In the limit of $\Lambda_{\rm NP} = M_{S}=M_{S'}=M_{\psi} = M_{\psi'}$ 
one finds:
\begin{align}
 I_{4}
 \rightarrow&
 \frac{1}{16\pi^{2}} \frac{1}{6} \frac{1}{\Lambda_{\rm NP}^{4}}.
\end{align}
The mean lifetime $\tau$ can then be estimated with the coefficient
of the effective operator, which gives:
\begin{align}
 \tau
 \simeq&
 \frac{1}{
 \frac{m_{p}}{32\pi}
 \left[
 1-\frac{m_{\pi^{+}}^{2}}{m_{p}^{2}}
 \right]^{2}
 \left|
 W_{0}
 (Y_{1})^{\alpha}
 (Y_{2})^{\beta}
 (Y_{3})^{\delta}
 (Y_{4})^{\gamma}
 M_{\psi} M_{\psi'} I_{4}
 \right|^{2}
 }
 \nonumber
 \\
\simeq &
 5 \cdot 10^{31} [\text{yrs}]
 \left|
 \frac{10^{-20}}
 {
 (Y_{1})^{\alpha}
 (Y_{2})^{\beta}
 (Y_{3})^{\delta}
 (Y_{4})^{\gamma}
 }
 \right|^{2}
 \hspace{-0.1cm}
 \left[
 \frac{
 \frac{1}{16 \pi^{2}}
 \frac{1}{6}
 \frac{1}{(3\text{ TeV})^{2}}
 }
 {M_{\psi} M_{\psi'} I_{4}}
 \right]^{2},
\end{align}
where we used the form factor $W_{0}$ given in
Ref.~\cite{Aoki:2013yxa}.  It turns out that the Yukawa couplings
$Y_{1,2,3,4}$ should be of the order $\lesssim\mathcal{O}(10^{-5})$ to
yield a proton decay signal detectable in the next generation
experiments, in case the new physics scale $\Lambda_{\rm NP}$ is of
the order of say a few TeV.\footnote{  We set
the new physics scale $\Lambda_{\rm NP}$ to be the TeV scale
to make our model testable at the LHC.
However larger values of $\Lambda_{\rm NP}$ are allowed,
which in turn implies larger values of the Yukawa couplings.}
Note that the experimental bound is
currently $\tau > 3.9\cdot 10^{32}$ yrs.

The coupling $Y_{3}$ is shared with the one-loop diagram of the Dirac
mass term for neutrinos, see Fig.~\ref{Fig:Diagrams}.  The neutrino
mass from this type of diagram has been calculated many times in the
literature. It can be written as:
\begin{align}
 {(m_{\nu})_{\alpha}}^{\beta}
 =
 \frac{s c (Y_{\nu}^{\dagger})_{\alpha}
 (Y_{3})^{\beta}M_{\psi'}
 }{16 \pi^{2} \sqrt{2}}
  \left[
 \frac{
 M_{\zeta_{1}}^{2}
 \ln \frac{ M_{\zeta_{1}}^{2} }{ M_{\psi'}^{2} }
 }
 { M_{\zeta_{1}}^{2} - M_{\psi'}^{2} }
 -
 \frac{
 M_{\zeta_{2}}^{2}
 \ln \frac{ M_{\zeta_{2}}^{2} }{ M_{\psi'}^{2} }
 }
 { M_{\zeta_{2}}^{2} - M_{\psi'}^{2} }
 \right],
\end{align}
where $s$ and $c$ are sine and cosine of the mixing angle between the
neutral components of the scalar mediators $\eta$ and $S'$ and their
mass eigenstates $\zeta_{1,2}$.  Assuming the magnitude of coupling
$\mu$ is set to be $\Lambda_{\rm NP}$, the same as all other
mass parameters, the size of the resulting Dirac neutrino mass is
estimated as
\begin{align}
 m_{\nu} \simeq&
 \frac{\langle H^{0} \rangle}{16\pi^{2}}
 Y_{\nu}^{\dagger} Y_{3}
 =
 {\cal O}(0.1) 
 \left[
 \frac{Y_{\nu}^{\dagger}}{10^{-5}}
 \right]
 \left[
 \frac{Y_{3}}{10^{-5}}
 \right]\text{[eV]}.
 \label{eq:mNu-estimation}
\end{align}
A more detailed fit to neutrino data could easily be
done~\cite{HHO:2018future}.
Interestingly, to have the correct size of neutrino masses and a
detectable rate for proton decay, keeping $\Lambda_{\rm NP}\sim$TeV,
the coupling $Y_{\nu}$ should also be roughly of order $Y_{\nu} \sim
\mathcal{O}(10^{-5})$.  Note that the connection
between the proton decay rate and the neutrino masses is not a one-to-one
correspondence.
This is because their diagrams only share the Yukawa coupling $Y_3$
(see Fig.~\ref{Fig:Diagrams}).

The Yukawa coupling $Y_{\nu}$ also mediates charged lepton flavour
violating (cLFV) processes, such as $\ell_{\alpha} \rightarrow
\ell_{\beta} + \gamma$ at the one-loop level. We estimate the decay rate 
with the general formulas given in Ref.~\cite{Lavoura:2003xp} as
\begin{align}
 \Gamma(\ell_{\beta} \rightarrow \ell_{\alpha} \gamma)
 \simeq
 \frac{e^{2}m_{\mu}^{5}}{16 \pi}
 \left|
 (Y_{\nu}^{\dagger})_{\alpha}
 (Y_{\nu})^{\beta}
 \left[
 -\bar{c} + \frac{3}{2} \bar{d}
 \right]
 \right|^{2},
\end{align}
where the loop integral $-\bar{c}+3\bar{d}/2$ is given
as a function of $t\equiv M_{\psi'}^{2}/M_{\eta^{+}}^{2}$;
\begin{align}
 -\bar{c} + \frac{3}{2} \bar{d}
 &=
 \frac{\rm i}{16 \pi^{2}}
 \frac{1}{M_{\eta^{+}}^{2}}
 \left[
 \frac{2t^{2} + 5t-1}{12(t-1)^{3}}
 -
 \frac{t^{2} \ln t}{2 (t-1)^{4}}
 \right]
 \nonumber
 \\
 &\xrightarrow[]{t\rightarrow 1}
 \frac{\rm i}{16 \pi^{2}}
 \frac{1}{24}
 \frac{1}{M_{\eta^{+}}^{2}}.
\end{align}
Using Yukawa couplings
$Y_{\nu}$ of order  $\mathcal{O}(10^{-5})$,
as suggested by neutrino masses (cf. Eq.~\eqref{eq:mNu-estimation}), 
and assuming the masses of the mediators are all at the TeV scale,
we find that the branching ratio for  $\mu \rightarrow e \gamma$
is roughly 
\begin{align}
 {\rm Br}(\mu \rightarrow e \gamma)
 =
 7\cdot 10^{-31}
 \left|
 \frac
 {Y_{\nu}^{\dagger} Y_{\nu}}{10^{-10}}
 \right|^{2}
 \left[
 \frac{3\text{ TeV}}{\Lambda_{\rm NP}}
 \right]^{4}.
\end{align}
This is far below current and future sensitivities 
\cite{Atanov:2018bfv}.  
In short, the correct order of neutrino masses can be reproduced, and
simultaneously the size of the signature mode $p \rightarrow
\pi^{+}+$missing of the light sterile neutrino can be kept at a detectable
size, while satisfying constraints from the cLFV and keeping the NP
scale stays at TeV. 

Finally, let us briefly mention dark matter and LHC phenomenology.  As
Tab.~\ref{Tab:Model} shows, the model has two neutral particles, one
fermion and one scalar. Both could be the dark matter (DM) depending
on which is the lightest state.  However, since
  both of our $Z_2$'s are broken softly, one would need to introduce
  another symmetry, to stabilize the DM candidate.  The simplest
  possibility is another $Z_2$, under which all beyond SM particles --
  except $N$ -- are odd. This symmetry also eliminates the terms 
in Eq.~(\ref{eq:Lagrangian2}).  Dark matter phenomenology for these
candidates has already been discussed in
Ref.~\cite{Farzan:2012sa}. Here we only note that our preferred
candidate  would be the neutral scalar, since the
fermionic candidate requires that $Y_{3}$ is much larger than
$10^{-5}$ in order to reproduce the correct relic density
\cite{Farzan:2012sa} and such large value of $Y_{3}$ would in turn
require quite a large hierarchy among the Yukawa couplings.

At the LHC, the new coloured particles in our model can be pair
produced through gluon-gluon fusion. Typical cross sections for the
scalars can be found in Ref.~\cite{Dorsner:2018ynv}. Cross sections for the
coloured fermions should be around a factor of two larger than those
for scalars (for the same mass).
The typical signature of the coloured particles are jet(s) with
missing energy.  The coloured scalar $S$ decays into a jet $(d_{R})$
with missing energy $\psi'$ through the Yukawa interaction $Y_{4}$.
The decay rate is roughly
\begin{align}
 \Gamma(S \rightarrow \overline{\psi'}_{L} + d_{R \alpha})
 =&
 \frac{3 |(Y_{4})^{\alpha}|^{2} M_{S}}{16 \pi}
 \left[
 1 - \frac{M_{\psi'}^{2}}{M_{S}^{2}}
 \right]^{2}.
\end{align}
With $M_{S} =$3 TeV and $Y_{4}=10^{-5}$, the decay rate is estimated
to be $2 \cdot 10^{-8}$ GeV (if $M_{S} \gg M_{\psi'}$), implying the
decay is prompt. $\overline{\psi'}_{L}$ will decay further, if it is not the
lightest neutral particle. However, $\overline{\psi'}_{L}$ decays invisibly,
thus there is no change in the LHC signature.
Leptoquark searches with the jet+$\bar{\nu}$ mode
at ATLAS \cite{Aaboud:2016qeg} and CMS \cite{CMS:2016imw} provide
currently lower limits on such coloured states, which are roughly of
the order of 1 TeV. However, all these searches are still based on
only moderate luminosity samples and significant improvements in these
searches in the high luminosity run of the LHC can be expected.
As an aside we note that if $Y_{4}$ is assumed to be much smaller, say
as small as $\mathcal{O}(10^{-9})$, the lifetime of $S$ becomes 
order of a nanosecond. The $S$ would then hadronize before decaying,
leaving an ionizing track in the detector, see for example the recent
paper \cite{Lenz:2016zpj} for a discussion of experimental status.


\section{Conclusions}

In this paper we have discussed a simple model that relates the 
longevity of the proton with the smallness of the neutrino mass. 
In this model, neutrinos are Dirac particles and proton decay 
is dominated by the final state $\pi^++\esl$. Although this is 
only an example model, we discussed at the level of effective 
$d=6$ operators, that in general the observation of proton decay 
with the final state  $\pi^++\esl$, together with the non-observation 
of the well-known $\pi^0e^+$ final state, could be interpreted 
in favour of the existence of a light sterile neutrino. 

We plan to study the details of the phenomenology of the model given in
this letter and exhaustively explore the relation between the proton
decay mode $p \rightarrow \pi^{+} +$missing and Dirac neutrino mass
models with the full decomposition of the proton decay operators 
$\mathcal{O}_{N1,N2}$~\cite{HHO:2018future}.

Finally, we would like to mention that the discussion based on the
effective operators, which is given in section \ref{sec:two}, is valid
also if the light sterile neutrino is not the Dirac partner of the
ordinary neutrino. Such a sterile neutrino could be an additional
Majorana neutrino, if $B-L$ is violated, or come with its own Dirac
partner otherwise.  Therefore, a positive result of sterile neutrino
searches in short baseline oscillation
experiments~\cite{Antonello:2015lea,Cianci:2017okw} would be
interesting, since it opens up the possibility for $d=6$ proton decay
operators to exist that exclusively produce the $\pi^{+}+\esl$ mode.

\begin{acknowledgements}

We thank Renato Fonseca for discussions on symmetries.
We thank Prof. Rabindra Mohapatra for
useful discussions on proton decay processes in GUT models.  This work
was supported by the Spanish MICINN grants FPA2017-85216-P,
SEV-2014-0398 and PROMETEOII/2014/084 (Generalitat
Valenciana). J.C.H. is supported by Chile grants Fondecyt No. 1161463,
Conicyt PIA/ACT 1406 and Basal FB0821.
\end{acknowledgements}


\end{document}